\documentclass[10pt,a4paper]{article}
\usepackage[vmargin=2.2 cm, hmargin = 2.2 cm]{geometry}
\usepackage{amsmath}
\usepackage{amssymb}
\usepackage{amsthm}
\usepackage{bm}
\usepackage{epsfig}
\usepackage{graphicx,xcolor}
\usepackage{multirow}
\usepackage{url}
\usepackage{algorithm}
\usepackage{algorithmic}
\usepackage{comment}

\newcommand{\MBI}{\mathbf{I}}

\newcommand{\SP}{\mathcal{P}}
\newcommand{\SF}{\mathcal{F}}

\newcommand{\WHD}{\widehat{D}}

\newcommand{\WHG}{\widehat{G}}
\newcommand{\WTlmda}{\widetilde{\lambda}}
\newcommand{\WDlmda}{\widehat{\lambda}}

\newtheorem{definition}{{}{\bf Definition}}[section]
\newtheorem{theorem}{{}{\bf Theorem}}[section]
\newtheorem{lemma}{{}{\bf Lemma}}[section]
\def\done{\hspace*{\fill} \rule{1.8mm}{2.5mm} \\}

\date{}

\begin{document}
\title{Mathematical Modeling of Insurance Mechanisms for E-commerce Systems}

\author{
Hong Xie  \,\,\,\,\, John C.S. Lui \\
Department of Computer Science \& Engineering \\
The Chinese University of Hong Kong\\
\{hxie,cslui\}\@cse.cuhk.edu.hk
}

\maketitle

\begin{abstract}
Electronic commerce (a.k.a. E-commerce)
systems such as eBay and Taobao of Alibaba
are becoming increasingly popular.
Having an effective reputation system
is critical to this type of internet service because
it can assist buyers to evaluate
the trustworthiness of sellers,
and it can also improve the revenue for
reputable sellers and E-commerce operators.
We formulate a stochastic model to
analyze an eBay-like reputation system and propose
four measures to quantify its effectiveness:
(1) new seller ramp up time,
(2) new seller drop out probability,
(3) long term profit gains for sellers, and
(4) average per seller transaction gains
   for the E-commerce operator.
Through our analysis, we identify key factors which
influence these four measures.
We propose a new insurance mechanism which
consists of an insurance protocol and a transaction mechanism to
improve the above four measures.
We show that our insurance mechanism
can reduce the ramp up time by around 87.2\%,
and guarantee new sellers ramp up before the deadline $T_w$
with a high probability (close to 1.0).
It also increases the long term profit gains
and average per seller transaction gains
by at least 95.3\%.
\end{abstract}

\section{{\bf Introduction}}

E-commerce systems are becoming increasingly popular and
typical examples include
eBay\cite{eBay}, Amazon\cite{Amazon},
and Taobao \cite{Taobao} of Alibaba
(the largest E-commerce system in China), etc.
Through an E-commerce system,
geographically distributed sellers and buyers can transact online.
Sellers advertise products in their online stores
(which reside in the E-commerce's website),
while buyers can
purchase products from any online stores.
The E-commerce system charges a transaction fee
from sellers for each completed transaction.
In an E-commerce system,
it is possible to purchase products from
a seller whom the buyer has never transacted with, and
this seller may not even be trustworthy\cite{Resnick2000}.
This situation results in a high risk of
buying low quality products.
To overcome such problems,
E-commerce systems deploy reputation
systems\cite{Resnick2000}.

Usually, E-commerce operators
maintain and operate a reputation mechanism
to reflect the {\em trustworthiness} of sellers\cite{eBay,Taobao}.
A high reputation seller can attract more transactions
leading to higher revenue\cite{Resnick2000}.
The eBay-like reputation system is
the most widely deployed reputation policy,
which is used in eBay and Tabao, etc.
This type of reputation system is a credit based system.
More precisely, a seller needs to collect enough
credits from buyers in order to improve his reputation.
These credits are obtained in form of feedback ratings,
which are expressed by buyers after each transaction.
Feedback ratings in eBay and Taobao are of three levels:
positive ($+1$), neutral ($0$), and negative ($-1$).
The cumulative sum of all the past feedback ratings
(i.e., reputation score)
reflects the {\em trustworthiness} of a seller.
The reputation score and feedback ratings
are public information and accessible by all
buyers and sellers in such an E-commerce system.

Consider this eBay-like reputation system,
a new seller may spend a long time to collect enough credits
(i.e., ramp up).
This is because new sellers are initialized with
a reputation score of zero, and buyers are less willing to
buy products from a seller with low reputation scores.
The ramp up time is critical to the effectiveness of
a reputation system.
A long ramp up time discourages new sellers to
join an E-commerce system.
Furthermore, a new user starts an online store with certain budgets,
and maintaining such online stores involves cost.
If a new seller uses up his entire budget
and has not yet ramped up his reputation,
he may discontinue his online business
(i.e, or drops out) due to low revenue.
Therefore, a long ramp up time increases the
risk that a new seller drops out and
discourages potential new sellers to join.
Finally, a long ramp up time also results in
a low profit gain for a seller.
Because before ramping up,
a seller can only attract few transactions
due to his low reputation score.
To an E-commerce operator, this
also results in an indirect loss on
transaction gains.
This paper aims to identify key factors
that influence the ramp up time and to design a
mechanism to improve this measure.

Reducing ramp up time is challenging and to the
best of our knowledge, this is the first work which explores
how to reduce the ramp up time for an eBay-like reputation system.
This paper aims to explore the following fundamental questions:
{\em
(1) How to identify key factors which influence the ramp up time?
(2) How to take advantage of these factors to reduce the ramp up time?
}
Our contributions are
\begin{itemize}
\item  We propose four performance measures to quantify
     the effectiveness of eBay reputation systems:
    (1) new seller ramp up time,
    (2) new seller drop out probability,
    (3) long term profit gains for sellers,  and
    (4) average per seller transaction gains
        for an E-commerce operator.

\item We develop a stochastic model to identify key factors
    which many influence these four measures.
    Through we gain important insights on how to design a new
    mechanism these performance measures.

\item We propose and design an insurance mechanism
    which can reduce
    the ramp up time and the new seller drop out probability.
    We show that our insurance mechanism
    can reduce the ramp up time by around 87.2\%,
    and guarantee new sellers ramp up before the deadline $T_w$
    with a high probability (close to 1.0).
    It also increases the long term profit gains
    and average per seller transaction gains
    by at least 95.3\%.

\end{itemize}

This paper organizes as follows.
In \S\ref{section: model}, we present the system model
  for E-commerce systems.
In \S\ref{section: problem} we formulate four measures to
  explore the ramp up time problem,
  i.e., ramp up time, new seller drop
  out probability, long term profit gains
  and average per seller transaction gains.
In \S\ref{section: reputation}, we derive
  analytical expressions for these four measures.
In \S\ref{section: insurance} present the design of our
  insurance mechanism.
Related work is given in \S\ref{section: relatedwork} and
  we conclude in \S\ref{section: conclusion}.

\section{{\bf E-commerce System Model}}
\label{section: model}

An E-commerce system consists of
{\em users}, {\em products} and a {\em reputation system}.
A user can be a {\em seller} or a  {\em buyer} or both.
Sellers advertise products in their online stores
and set a price for each product.
Buyers, on the other hand,
purchase products through online stores
and provide feedbacks to indicate whether
a buyer advertises products honestly or not.
A reputation system is maintained by E-commerce operators
to reflect the {\em trustworthiness} of sellers.
A high reputation seller can attract more transactions
leading to a high revenue.
The reputation system aggregates all the feedbacks,
and computes a reputation score for each seller.
The reputation score is public information
which is accessible by all buyers and sellers.

Products are categorized into different types.
For example, eBay categorizes products into
``Fashion'', ``Electronics'', ``Collectibles \& Art'',
etc~\cite{eBay}.
We consider $L \geq 1$ types of product.
Consider a type $\ell \in \{1, \ldots, L\}$ product.
A seller sets a price $p_\ell \in [0,1]$
and the E-commerce operator charges a transaction fee of
$T \triangleq \alpha p_\ell$, where $\alpha \in (0,1)$,
after the product is
sold\footnote{We can also consider a fixed transaction fee model
and our analysis is still applicable. But for brevity, let us consider
a transaction fee which is proportional to the selling price.}.
It has a manufacturing cost of $c_\ell \!\in\! [0,1]$.
A seller earns a profit of $u_\ell$ by selling one product, we have
\begin{equation}
 u_\ell \!=\! (1 - \alpha) p_\ell - c_\ell.
\label{equation: unit profit}
\end{equation}
For the ease of presentation,
our analysis focuses on one product type.
It can be easily generalized to multiple
product types,
and we omit the subscript unless we state otherwise.

\subsection{{\bf Transaction Model}}

Sellers advertise the product quality in their online stores.
Let $Q_a\!\in\![0,1]$ be the
{\em advertised quality}.
The larger the value of $Q_a$ implies
the higher the advertised quality.
Buyers refer to the advertised
quality $Q_a$ in their product adoption.
Each online store also has an intrinsic quality.
Let $Q_i \!\in\![0,1]$ be the {\em intrinsic quality}
(i.e., the ground truth of the product's quality).
The larger the value of $Q_i$ implies
the higher the intrinsic quality.
Since sellers aim to promote their products,
so we have $Q_a \!\geq\! Q_i$.
We emphasize that the intrinsic quality $Q_i$
is private information, e.g., it is only known to the seller.
On the other hand, the advertised quality $Q_a$
is public information which is accessible by
all buyers and sellers.

Buyers estimate the product quality by referring to
the advertised quality $Q_a$
(we will present the estimating model later).
Let $Q_e \!\in\! [0,1]$ be the {\em estimated quality}.
The larger the value of $Q_e$ implies
the higher the estimated quality.
To purchase a product,
a buyer must submit a payment $p$ to the E-commerce system,
which will be given to the corresponding
seller when he receives the product.
There is usually a shipment delay in any E-commerce systems.
We denote the delay as $d$.
Upon receiving a product,
a buyer can evaluate its quality and at that moment,
he has the {\em perceived quality}, which we denote as
$Q_p \!\in\![0,1]$.
The larger the value of $Q_p$ implies
the higher the perceived quality.
We assume that buyers can perceive
the {\em intrinsic quality}, i.e., $Q_p \!=\! Q_i$.
Buyers are satisfied (disappointed)
if they find out that the product is
at least as good as (less than) it is advertised, or
$Q_p \!\geq\! Q_a$ ($Q_p < Q_a$).

To attract buyers,
an E-commerce system needs to incentivize
sellers to advertise honestly, i.e., $Q_a \!=\! Q_i$.
Many E-commerce systems achieve this by
deploying a reputation system.
We next introduce a popular reputation
system used by many E-commerce systems
such as eBay~\cite{eBay} or Taobao~\cite{Taobao}.
Table~\ref{Tb:Model:Notations} summarizes key notations
in this paper.
\begin{table}[htb]
\centering
\begin{tabular}{|c|l|} \hline
$p, c$
  & price and  manufacturing cost of a product    \\ \hline
$T, u$
  & transaction fee, unit profit of selling a product \\ \hline
$Q_a, Q_i, Q_e, Q_p$
  & advertised, intrinsic, estimated, perceived product quality \\ \hline
$d, C_S$
  & shipment delay, shipment cost \\ \hline
$\gamma$
  & critical factor in expressing feedback ratings \\ \hline
$\SF$
  & reputation profile for a seller \\ \hline
$r$
  & reputation score for a seller \\ \hline
$r_h, \theta$
  & reputation threshold, consistency threshold \\ \hline
$\beta$
  & discounting factor in estimating product quality \\ \hline
\multirow{2}{*}{$\SP(Q_e, p)$}
  & probability that a buyer buys a product with  \\
  & an estimated quality quality $Q_e$ and a price $p$ \\ \hline
\multirow{2}{*}{$P_{ba}, P_{br}$}
  & probability that a buyer buys a product from a seller  \\
  & labelled as average (reputable)\\ \hline
$\lambda_1 (\lambda_2)$
  & buyer's arrival rate before (after) a seller ramps up \\ \hline
\multirow{2}{*}{$T_w$}
  & the maximum time that a seller\\
  & is willing to wait to get ramped up \\ \hline
$T_r, P_d$
  & ramp up time, new seller drop out probability \\ \hline
\multirow{2}{*}{$G_s$, $G_e$}
  & long term expected profit gains for a seller, average per  \\
  & seller transaction gains for the E-commerce operator\\ \hline
$\delta$
  & discount factor in the long term expected profit gains $G_s$\\ \hline
$\lambda_T (\tau)$
  & transaction's arrival rate at time slot $\tau$\\ \hline
$C_I, D_I, T_d, T_c$
  & insurance price, deposit, duration time and clearing time \\ \hline
$\WHD_I$
  & insurance deposit threshold to revoke insurance certificate\\ \hline
$\lambda_I$
  & transaction's arrival rate to an insured seller \\ \hline
\multirow{2}{*}{$T_r^I$, $P_r^I$}
  & ramp up time with insurance,  \\
  & new seller drop out probability with insurance \\ \hline
\multirow{2}{*}{$G_s^I, G_e^I$}
  & long term expected profit gains with insurance,  \\
  & average per seller transaction gains with insurance \\\cline{2-2} \hline
\end{tabular}
\vspace{0.05 in}
\caption{Notation list}
\label{Tb:Model:Notations}
\end{table}

\subsection{{\bf Baseline Reputation System}}

The eBay-like system maintains a reputation system
to reflect the trustworthiness of sellers.
It consists of a {\em feedback rating system} and
{\em a rating aggregation policy}.

Buyers express feedback ratings to indicate
whether a seller advertises honestly or not.
The eBay-like system adopts a feedback rating system consisting
of three rating points\footnote{We can easily generalize
the model to consider more rating points.}, i.e., $\{-1, 0, 1\}$.
A positive rating (rating $1$) indicates that
a product is at least as good as it is advertised,
i.e., $Q_p \geq Q_a$.
A neutral rating (rating $0$) indicates that a buyer
is indifferent about the product that he purchased.
This happens when the perceived quality is slightly
lower than it is advertised,
i.e., $Q_p \!\in\! [Q_a - \gamma, Q_a)$,
where $\gamma \!\in\! [0,1]$ denotes the critical factor.
The smaller the value of $\gamma$ implies
that buyers are more critical in expressing ratings,
e.g., $\gamma \!=\! 0$ means that
buyers have zero tolerance on seller overstating
the product quality.
A negative rating (rating $-1$) represents that
the perceived quality is
far smaller than the advertised quality,
i.e., $Q_p \!<\! Q_a - \gamma$.
We have
\[
\mbox{feedback rating} = \left\{
\begin{aligned}
& 1,
  && \mbox{if $Q_p \geq Q_a$},  \\
& 0,
  && \mbox{if $Q_a - \gamma \leq Q_p < Q_a$}, \\
& -1,
  && \mbox{if $Q_p < Q_a - \gamma$}.
\end{aligned}
\right.
\]
All the historical ratings are known to
all buyers and sellers.

\begin{figure}[htb]
\centering
\includegraphics[width=0.35\textwidth]{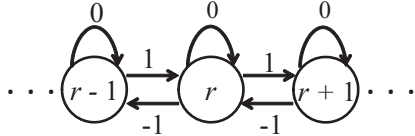}
\caption{Transition diagram of a seller's reputation score $r$ }
\label{Model:reputation}
\end{figure}

For the rating aggregation policy, each
seller is associated with a reputation score,
which is the summation of all his feedback ratings.
We denote it by $r \in \mathbb{Z}$.
A new seller who enters the E-commerce system
is initialized with zero reputation score, or $r = 0$.
A positive feedback rating increases $r$ by one,
a negative feedback rating decreases $r$ by one,
and a neutral feedback rating $0$ does not change $r$.
Figure~\ref{Model:reputation} depicts the
transition diagram of a seller's reputation score.
Note that $r$ is a public
information accessible by all buyers and sellers.

To assist buyers to evaluate the trustworthiness
of a seller,
E-commerce systems not only announce the
seller's reputation score $r$,
but also his reputation profile.
Let $\SF \!\triangleq\! (r, n^+, n^0, n^-)$
be the reputation profile,
where $n^+, n^0, n^-$ represent the {\em cumulative number of
feedback ratings} equal to $1, 0, -1$ respectively.
Note that this form of reputation is commonly deployed,
say in eBay~\cite{eBay} and Taobao~\cite{Taobao}.

Shipment delay in real-world E-commerce systems usually
results in certain delay in the reputation update.
To characterize the dynamics
of a reputation updating process,
we consider a discrete time system
and divide the time into slots,
i.e., $[0,d), [d,2d), \ldots$,
where $d$ is the shipment delay.
We refer to a time slot $\tau \!\in\! \mathbb{N}$
as $[\tau d, (\tau+1)d)$.
Let $N (\tau)$ be the number of products
sold in the time slot $\tau$.
Suppose $N^+ (\tau), N^0 (\tau), N^- (\tau)$
of these transactions result in
positive, neutral and negative feedbacks respectively.
Let
$\SF (\tau) \!\triangleq\!
(r(\tau), n^+(\tau), n^0(\tau), n^-(\tau))$
be the reputation profile at time slot $\tau$.
Then we have $\SF(0) \!=\!(0,0,0,0)$.
We update the reputation profile $\SF(\tau)$ as
\begin{equation}
\left\{
\begin{aligned}
& n^+(\tau+1) = n^+(\tau) + N^+(\tau), \\
& n^0(\tau+1) = n^0(\tau) + N^0(\tau),  \\
& n^-(\tau+1) = n^-(\tau) + N^-(\tau), \\
& r(\tau + 1) = r(\tau) + N^+(\tau) -  N^-(\tau).
\end{aligned}
\right.
\label{Model:Eq:reputationUpdating}
\end{equation}
For simplicity, we drop the time stamp $\tau$ in
the reputation profile, when no confusion involved.

We next present a model to characterize
the impact of sellers' reputation profiles on
buyers' product adoption behavior.
This model serves as an important building
block to explore the effectiveness of
this baseline reputation system.

\subsection{{\bf Model for Product Adoption Behavior}}

The reputation system forges trust among
sellers and buyers.
This trust plays a critical role in product adoption.
More precisely, buyers evaluate the trustworthiness
of sellers from sellers' reputation profiles.
Buyers seek to minimize the risk in product purchase
and they prefer to buy from reputable sellers.

Based on the reputation profile $\SF$,
our model classifies sellers into two types:
``{\em reputable}'' and ``{\em average}''.
To be labeled as reputable,
a seller's reputation profile must satisfy two conditions.
The first one is that a seller needs to collect enough credits,
i.e., positive feedbacks from buyers.
More precisely, his reputation score must be
at least greater than or equal to
some positive reputation threshold $r_h$,
i.e., $r \!\geq\! r_h$.
A new seller is initialized with zero reputation score,
i.e., $r \!=\! 0$.
To accumulate a reputation score of at least $r_h$,
he needs to accomplish enough number of honest transactions.
The second condition is that a seller should
be {\em consistently honest}.
More concretely, the fraction of positive feedbacks
should be larger than or equal to a consistency threshold
$\theta \!\in\!(0,1]$,
i.e., $n^+ / (n^+ + n^- + n^0) \geq \theta$.
The larger the value of $\theta$ implies that the E-Commerce operators
are more critical about the honest consistency.
We formally define a reputable seller
and an average seller as follows.
\begin{definition}
A seller is labeled as reputable if and only if the follwing two
conditions are met
\begin{enumerate}
\item[C1:]   $r \geq r_h$ and,
\item[C2:] $n^+ / (n^+ + n^- + n^0) \geq \theta$.
\end{enumerate}
Otherwise, a seller is labeled as an average seller.
\end{definition}
\noindent
Hence, the reputation threshold $r_h$
and consistency  threshold $\theta$ quantify how difficult is
it to earn a reputable label.
The larger the $r_h$ and $\theta$,
the more difficult it is to earn a reputable label.

A buyer estimates the product quality referring to
the advertised quality $Q_a$ and
the reputation profile of a seller.
More concretely, if a seller's reputation profile
indicates that this seller is reputable,
then a buyer believes that this seller advertises honestly.
This buyer therefore estimates the product
quality as the advertised quality, i.e., $Q_e \!=\! Q_a$.
On the contrary, if the reputation profile indicates
that a seller is average,
a buyer believes that this seller is likely to overstate
the product quality.
Hence the estimated quality is lower
than the advertised quality, i.e., $Q_e \!=\! \beta Q_a$,
where $\beta \!\in\![0,1]$
denotes the discounting factor.
The smaller the value of $\beta$ implies that buyers are
less willing to trust an average seller.
We have
\[
Q_e = \left\{
\begin{aligned}
& Q_a,
  && \mbox{if $r \geq r_h$ and
     $n^+ / (n^+ + n^- + n^0) \geq \theta$,} \\
& \beta Q_a,
  && \mbox{otherwise.}
\end{aligned}
\right.
\]

A buyer makes the purchasing decision based on the
estimated quality $Q_e$ and the product price $p$.
More concretely, the probability that a buyer buys a product
increases in $Q_e$
and decreases in $p$.
Formally, we have
\begin{equation}
\Pr[\mbox{adopts a product}] \triangleq
\SP\left(Q_e, p\right),
\label{Model:eq:AdoptProb}
\end{equation}
where the function $\SP$ increases
in $Q_e$ and decreases in $p$.

In the following section,
we evaluate the effectiveness of the baseline
reputation system in E-commerce applications.
Our goal is to identify key factors that
influence the effectiveness of this reputation system and if possible,
improve it.

\section{{\bf Problems Formulation}} \label{section: problem}

We propose four performance measures
to quantify the effectiveness of the
baseline reputation system mentioned in
Section~\ref{section: model}.
These measures are:
(1) ramp up time $T_r$,
(2) new seller drop out probability $P_d$,
(3) long term expected profit gains for a seller $G_s$, and
(4) average per seller transaction gains for
    the E-commerce system operator $G_e$.
We also present our problem formulations and
our objective is to identify key factors
which can influence these measures.
Lastly, we raise an interesting question of
whether there are other mechanisms which can
reduce the ramp up time and
the new seller drop out probability,
and improve the long term expected profit gains
and average per seller transaction gains.

\subsection{{\bf Ramp Up Time}}

Sellers and E-commerce system operators are
interested in the minimum time that a new seller
must spend to collect enough credits,
i.e., positive feedbacks from buyers,
so that the seller can be classified as reputable.
For one thing, a reputable seller can attract more buyers
which may result in more transactions,
and higher transaction volume implies
higher transaction gains to the E-commerce operator.
We next formally define the ramp up process and the
ramp up condition.

\begin{definition}
A new seller's reputation is $r \!=\! 0$.
He needs to collect enough credits,
i.e., positive feedbacks from buyers,
so that his reputation $r$ can increase to at least $r_h$.
The process of increasing his reputation to $r_h$ is called
the ramp up process.
Furthermore, when $r \!\geq\! r_h$, then we say
that the ramp up condition is satisfied.
\end{definition}

Recall that $r(\tau)$ denotes the reputation score of
a seller at time slot $\tau$.
We formally define the ramp up time as follows.
\begin{definition}
Ramp up time is the minimum time
that a seller must spend to accumulate a reputation score of $r_h$.
Let $T_r$ denote the ramp up time, we have
\begin{equation}
T_r \triangleq d \cdot \arg\min_\tau \{ r(\tau) \geq r_h\}.
\label{Problem:Eq:RampTime}
\end{equation}
\end{definition}
\noindent
The ramp up time quantifies how long it will take
to collect enough credits from buyers.
It is critical to the profit gains for a seller.
To see this, we next
quantify how the ramp up time can affect
the transaction's arrival rate.

A seller can attract more buyers
when he satisfies the ramp up condition because
his online store will receive higher click rate by buyers,
therefore increasing his profit gains.
Let $\lambda_1$ ($\lambda_2$) be the buyer's arrival rate
before (after) a seller satisfies the ramp up condition.
We assume that the buyer's arrival process,
either before or after a seller satisfies the ramp up condition,
follows a Poisson counting process
with parameter $\lambda_1$ (before ramping up)
and $\lambda_2$ (after ramping up) respectively,
where $\lambda_1 \!<\! \lambda_2$ to signify
that a ramped up seller can attract more buyers.
Recall that in Equation~(\ref{Model:eq:AdoptProb})
we express the probability that a buyer
adopts a product as $\SP(Q_e, p)$.
If a buyer adopts a product,
we say a seller obtains a transaction.
Based on the Poisson property, it is easy to see that the
transaction's arrival process is also
a Poisson counting process.
Let $\lambda_T(\tau)$ be the transaction's arrival rate
at time slot $\tau$.
Let $\SP(Q_e(\tau), p)$ be the probability that
a buyer adopts a product at time slot $\tau$,
where $Q_e (\tau)$ denotes the estimated quality
at time slot $\tau$.
We can express the transaction's arrival rate as
\begin{align}
\lambda_T(\tau) =
& \left\{
\begin{aligned}
& \lambda_1 \SP(Q_e(\tau), p),
  && \mbox{if $r(\tau) < r_h$, } \\
& \lambda_2 \SP(Q_e(\tau), p),
  && \mbox{if $r(\tau) \!\geq\! r_h$,}
\end{aligned}
\right. .
\label{Model:TransactionRate}
\end{align}
Equation~(\ref{Model:TransactionRate}) serves
as an important building block for us to
explore the key factors which influence the ramp up time $T_r$.
Let us formulate our first problem.

\noindent
{\bf Problem 1:}
{\em Identify key factors
which influence the ramp up time $T_r$,
and design a mechanism which can take advantage
of these factors to reduce $T_r$.
}

\subsection{{\bf New Seller Drop Out Probability}}

In real-world E-commerce systems,
a new seller may drop out,
or move to another E-commerce system,
if he does not collect enough credits
(i.e., ramp up) within certain time
because he cannot obtain enough transactions.
For example, a new seller in eBay may drop out
if he does not ramp up in one year.
This is because a new seller starts
an online store with certain budgets
and there are costs associated with maintaining
this online business.
Let $T_w > 0$ denote the maximum time that
a new seller is willing to wait to get ramped up.
In other words, if the ramp up time is longer than $T_w$,
a new seller will quit or drop out from that E-commerce system.
We assume $T_w /d \in \mathbb{N}$ to accommodate the
delay ($d$) in reputation update.
\begin{definition}
A new seller drops out, if and only if $T_r \!>\! T_w$.
\end{definition}

Sellers and the E-commerce operator are interested
in this new seller drop out probability.
Let
\begin{equation}
P_{d} \triangleq \Pr[T_r > T_w]
\label{Problem:Eq:DropProb}
\end{equation}
denote the probability that a new seller drops out.
The smaller the value of $P_{d}$ implies that sellers are
more likely to continue his online business in the E-commerce system.
This is an important measure since a small $P_{d}$ can attract more
new sellers to join the E-commerce system,
which will result in higher transaction gains
for the E-commerce operator.
On the other hand, a large $P_d$ discourages new sellers
to participate and can reduce the transaction
gains for the E-commerce operator.
We therefore consider the second problem.

\noindent
{\bf Problem 2:}
{\em Identify key factors
which influence the new seller drop out probability $P_d$,
and design a mechanism which can take advantage
of these factors to reduce $P_d$.
}

\subsection{{\bf Long Term Profit Gains and Transaction Gains}}

The profit gain (transaction gain) is critical to
sellers (E-commerce system operators).
We focus on the scenario that sellers
are long lived and they aim to
maximize their long term profit gains.
Recall that $u$, the unit profit of selling one product,
is expressed in Equation (\ref{equation: unit profit}).
Also recall that $N (\tau)$ denotes the number of products
sold in the time slot $\tau$.
We emphasize that $N(\tau)$ is a random variable and
follows a Poisson distribution
with parameter  $\lambda_T (\tau) d$,
where $\lambda_T (\tau)$ is derived in
Equation~(\ref{Model:TransactionRate}).
A seller earns a profit of $u N (\tau)$
in the time slot $\tau$.
We consider a discounted long term profit gain with
a discounting factor of $\delta \in (0, 1]$.
Let $G_s$ denote
the long term expected profit gains for a seller.
We express it as
\begin{equation}
G_s \triangleq E \left[
\sum_{\tau = 0}^\infty \delta^{\tau} u N (\tau)
\right].
\label{equation: profit gain}
\end{equation}
Note that when a seller earns a profit $u$,
he also contributes a transaction fee $T=\alpha p$
to the E-commerce operator.
Let $G_e$ denote the average per seller transaction gains
that a seller pays to the E-commerce operator.
We can express it as
\begin{equation}
G_e \!\triangleq\!
E \left[ \sum_{\tau = 0}^\infty \delta^{\tau} T N (\tau) \right] \!=\!
E \left[ \sum_{\tau = 0}^\infty \delta^{\tau} \alpha p N (\tau) \right] \!=\!
\frac{\alpha p}{u} G_s
.
\end{equation}
Note that $G_s$ is important to a seller while $G_e$ is important
to the E-commerce operator.
We consider the following problem.

\noindent
{\bf Problem 3:}
{\em
Identify key factors
which influence the profit gains
$G_s$ and average per seller transaction gains $G_e$,
and design a mechanism to use
these factors to improve $G_e$ and $G_s$.
}

We next derive $E[T_r]$, $P_d$, $G_s$, and $G_e$.
Through this analysis, we identify key factors which influence
the above mentioned performance measures.
These insights will serve as important building blocks for us
to design a mechanism.

\section{{\bf Analyzing the Baseline Reputation System}}
\label{section: reputation}

We derive analytical expressions for
the expected ramp up time ($E[T_r]$),
the new seller drop out probability ($P_d$),
the long term expected profit gains ($G_e$)
and the average per seller transaction gains ($G_e$).
Through this we identify that
the reputation threshold ($r_h$), as well as
the probability that a buyer buys a product from
an ``average labeled'' seller ($P_{ab}$)
are two critical factors which influence
$T_r$, $P_d$, $G_s$ and $G_e$.
Our results indicates that the baseline reputation mechanism
described in Section \ref{section: model},
suffers from long ramp up time,
high new seller drop out probability,
and small long term profit gains or transaction gains.
These insights show that one need to have
a new mechanism to reduce $T_r$, $P_d$,
and to improve $G_s$ and $G_e$.
We will present this new mechanism in
Section~\ref{section: insurance}.

\subsection{{\bf Deriving the Expected Ramp Up Time $E[T_r]$}}

Let us derive the analytical expression for
the expected ramp up time $E[T_r]$.
This measure quantifies on average, how long it will
take to ramp up a new seller under the baseline
reputation mechanism mentioned in Section~\ref{section: model}.
We consider the scenario that
buyers advertise the product quality honestly,
i.e., $Q_a \!=\! Q_i$.
As to how an eBay-like reputation mechanism can guarantee
rational sellers to advertise honestly,
one can refer to~\cite{Dellarocas2001}.
We like to point out that new sellers can achieve the {\em lowest
ramp up time} by advertising honestly ($Q_a \!=\! Q_i$).
Hence, the assumption that $Q_a \!=\! Q_i$
can be viewed as deriving the best case of $T_r$
for the baseline reputation system.
Let us first define the following notations.
\begin{definition}
Let $P_{ba} \!\triangleq\! \SP (\beta Q_i, p)$ and
$P_{br} \!\triangleq\! \SP (Q_i, p)$
denote the probability that a buyer buys a product
from an ``average labeled'' seller and a ``reputable''
seller respectively.
\end{definition}

\noindent
In the following theorem, we state the expected ramp up time.

\begin{theorem}
The expected ramp up time is
\[
E[T_r] = d \sum_{\tau=1}^\infty
\left(1 - \sum_{k=r_h}^{\infty}
e^{- \lambda_1 P_{ba} (\tau-1) d}
\frac{(\lambda_1 P_{ba} (\tau -1 ) d )^k}{k!}
\right).
\]
The $E[T_r]$ increases in the reputation threshold $r_h$,
and decreases in the transaction's arrival rate
$\lambda_1 P_{ba}$.
\label{Reputation:Thm:Rampuptime}
\end{theorem}
\noindent
{\bf Proof:}
Please refer to the appendix for derivation.
\done

\noindent
{\bf Remark:}
Theorem~\ref{Reputation:Thm:Rampuptime} states
that a new seller is more difficult to get ramped up
if the E-Commerce operator sets a high reputation threshold $r_h$,
or the transaction's arrival rate
to an ``average labeled'' seller
($\lambda_1 P_{ba}$) is low.

Table~\ref{Tb:Reputation:RampUpTime} presents numerical
examples on the expected ramp up time $E[T_r]$,
where we fix $P_{ba}\!=\!0.02$, i.e.,
buyers buy products from an ``average labeled'' seller
with probability 0.02,
and fix $d \!=\! 3$,
i.e., it takes three days to ship a product to a buyer
(or for the E-commerce operator to update
sellers' reputation).
We vary the buyer's arrival rate $\lambda_1$
from 5 to 25, i.e.,
on average each day an ``average labeled'' seller
attracts 5 to 25 buyers to visit his online store.
We vary the reputation threshold ($r_h$) from 100 to 200.
When $r_h \!=\! 200$,
as $\lambda_1$ increases from 5 to 25,
the expected ramp up time ($E[T_r]$) drops from
2001.7 to 401.6 days, a deduction ratio of $80\%$.
When the buyer's arrival rate is low,
say $\lambda_1 \!=\! 5$,
as the reputation threshold $r_h$
drops from 200 to 100,
the expected ramp up time $E[T_r]$ drops from
2001.7 to 1001.4 days, a reduction ratio of $50\%$.
These results indicates that
the expected ramp up time ($E[T_r]$)
is large in general.
Namely, it is difficult for new sellers to
quickly get ramped up under the baseline reputation system.
We next explore the new seller drop out probability.

\begin{table}[htb]
\centering
\begin{tabular}{|c|c|c|c|c|c|}
\hline
$\lambda_1$ & 5   & 10   & 15   & 20 & 25 \\ \hline \hline
$E[T_r]$ ($r_h \!=\! 200$)
  & 2001.7 & 1001.4 & 668.2 & 501.5 & 401.6 \\ \hline
$E[T_r]$ ($r_h \!=\! 150$)
  & 1501.4 & 751.5  & 501.5 & 376.5 & 301.5  \\ \hline
$E[T_r]$ ($r_h \!=\! 100$)
  & 1001.4 & 501.5  & 334.8 & 251.5 & 201.5 \\ \hline
\end{tabular}
\vspace{0.05 in}
\caption{Expected ramp up time $E[T_r]$ in days
  ($P_{ba} = 0.02, d\!=\! 3$).}
\label{Tb:Reputation:RampUpTime}
\end{table}

\subsection{{\bf Deriving the New Seller Drop Out Probability $P_d$}}

We now derive the analytical expression for $P_d$.
This probability quantifies how difficult it
is for a new seller to {\em survive} in the E-commerce system.
Note that $P_d$ is also crucial for new
sellers to decide whether or not to open an
online store in an E-commerce system.
Namely, a low drop out probability $P_d$
is attractive to new sellers,
while a high $P_d$ discourages new sellers to join.

\begin{theorem}
The new seller drop out probability is
\[
P_d = \sum_{k=0}^{r_h - 1}
e^{- \lambda_1 P_{ba} T_w }
\frac{(\lambda_1 P_{ba} T_w )^k}{k!}.
\]
The $P_d$ decreases in $\lambda_1P_{ba}$, $T_w$
and increases in $r_h$.
\label{Reputation:Thm:BankruptProb}
\end{theorem}
\noindent
{\bf Proof:}
Please refer to the appendix for derivation.
\done

\noindent
{\bf Remark:}
Theorem~\ref{Reputation:Thm:BankruptProb}
states that a new seller can reduces the drop out
probability by extending his ramp up deadline line ($T_w$),
and a new seller is more likely to
drop out if the reputation threshold ($r_h$) increases
or the transaction's arrival rate
to an ``average labeled'' seller
($\lambda_1 P_{ba}$) decreases.

Table~\ref{Tb:Reputation:BankruptProb} presents
numerical examples on the new seller drop out
probability $P_d$,
where we set $\lambda_1 = 20$,
i.e., on average, each day an ``average labeled'' seller
attracts 20 buyers to visit his store,
$d \!=\! 3$, and $T_w \!=\! 180$, i.e.,
sellers drops out if they do not ramp up in 180 days.
We vary $P_{ba}$,
the probability that a buyer buys products from an
``average labeled'' seller,
from 0.01 to 0.05,
and vary the reputation threshold $r_h$
from 100 to 200.
Consider $r_h \!=\! 200$.
As $P_{ba}$ increases from 0.01 to 0.05,
the new seller drop out probability $P_d$ decreases
from 1 to 0.92514.
This implies a very high drop out probability.
Consider $P_{ba} \!=\! 0.03$.
As the reputation threshold $r_h$ drops from 200 to 100,
we see that $P_d$ drops from 1 to 0.20819,
a reduction ratio of around $80\%$.
It is interesting to observe that when the $P_{ba}$
is small, the new seller drop out probability is quite high.
In fact when $P_{ba} \!=\! 0.01$, $P_d$
is very close to 1.
In other words, if buyers are less willing to
buy from ``average labeled'' sellers,
new sellers will be more likely to drop out.
We next explore key factors which influence
long term expected profit gains and
average per seller transaction gains.

\begin{table}[htb]
\centering
\begin{tabular}{|c|c|c|c|c|c|}
\hline
$P_{ba}$ & 0.01   & 0.02   & 0.03   & 0.04 & 0.05 \\ \hline \hline
$P_d$ ($r_h=200$) & 1.00000 & 1.00000 & 1.00000 & 0.99999 & 0.92514 \\ \hline
$P_d$ ($r_h=150$) & 1.00000 & 1.00000 & 0.99992 & 0.68056 & 0.00991 \\ \hline
$P_d$ ($r_h=100$) & 1.00000 & 0.99897 & 0.20819 & 0.00005 & 0.00000 \\ \hline
\end{tabular}
\vspace{0.05 in}
\caption{New seller drop out probability $P_d$
  ($\lambda_1 = 20, T_w \!=\! 180, d \!=\! 3$).}
\label{Tb:Reputation:BankruptProb}
\end{table}

\subsection{{\bf Deriving the Long Term Profit Gains $G_s$ and $G_e$}}

Let us now derive analytical expressions
for the long term expected profit gains
$G_s$ and the average per seller transaction gains $G_e$
respectively.
They are important measures
because a large $G_s$ is attractive to new sellers
and a small $G_s$ discourages new sellers to
join the E-commerce system,
while the average per seller transaction gains
$G_e$ is crucial to the E-commerce system operator.

\begin{theorem}
The long term expected profit gains for a new seller
can be expressed as
\begin{align*}
& G_s
  =
  \sum_{k = 0}^{r_h -1}
  \frac{(\lambda_1 P_{ba} T_w)^{k}}
    {(k)!}
  e^{- \lambda_1 P_{ba} T_w}
  \int \ldots \int\limits_{
  \hspace{-0.35 in}
  0< t_1 < \ldots < t_k < T_w }
  \frac{k!}{T_w^{k}}
  \sum_{j=1}^k u\\
&
  \delta^{\lceil t_j / d \rceil}
  dt_1 \ldots dt_k
  +
  \int_0^{T_w}
  \!\!dt_{r_h}\!\!\!
  \int \ldots \int\limits_{
  \hspace{-0.4 in}
  0< t_1 < \ldots < t_{r_h -1} < t_{r_h}}
  \!\!\!\!
  \lambda_1 P_{ba}
  e^{- \lambda_1 P_{ba} t_{r_h}}\\
&
  \frac{(\lambda_1 P_{ba} t_{r_h})^{r_h - 1}}
    {(r_h - 1)!}
  \frac{(r_h - 1)!}{(t_{r_h})^{r_h -1}}
  u
  \bigg(\!
  \sum_{j=1}^{r_h} \delta^{\lceil t_j/d \rceil}
  \!+\!
  \lambda_2 P_{br} d
  \frac{\delta^{\lceil t_{r_h}/d \rceil + 1} }{1 - \delta}
   \\
&
  +
  \delta^{\lceil t_{r_h}/d \rceil} \lambda_1 P_{ba}
  (d \lceil t_{r_h} / d \rceil - t_{r_h})
  \bigg)
  dt_1 \ldots dt_{r_h - 1}.
\end{align*}
Furthermore,
$G_e = \frac{\alpha p}{u} G_s$.
\label{Reputation:Thm:Revenue}
\end{theorem}
\noindent
{\bf Proof:}
Please refer to the appendix for derivation.
\done

\noindent
{\bf Remark:}
Theorem~\ref{Reputation:Thm:Revenue}
quantifies the impact of various factors on the
long term expected profit gains ($G_s$) and
average per seller transaction gains ($G_e$),
e.g., the reputation threshold $r_h$,
the buyer's arrival rate $\lambda_1, \lambda_2$, etc.
However the analytical expressions are in complicated forms.
It is hard to draw some insights on how
critical are these factors in influencing $G_s$ and $G_e$
by directly examining these expressions.
We next present some numerical examples to
illustrate their impact.

Table~\ref{Tb:Reputation:Revenue} presents numerical
examples on the long term expected profit gains ($G_s$)
and the average per seller transaction gains ($G_e$),
where we set $\lambda_1 \!=\! 20$, $T \!=\! \alpha p \!=\! 0.1$,
$\lambda_2 \!=\! 50$, $u \!=\! 1$,
$T_w \!=\! 180$, $d \!=\! 3$, $P_{br} \!=\! 0.1$
(i.e., buyers buy products from
a reputable seller with probability 0.1).
We vary $P_{ba}$, the probability that a buyer purchases
products from an ``average labeled'' seller,
and the ramp up threshold $r_h$ to
examine their impact on $G_s$ and $G_e$.
Consider $r_h \!=\! 200$.
As $P_{ba}$ increases from 0.01 to 0.05,
$G_s$ improves from 26.833 to 198.059,
an improvement ratio of 6.38 time.
Similarly, the average per seller transaction gains $G_e$
is also improved by 6.38 times.
This implies that $P_{ba}$
is critical to both sellers' profit gains and
the E-commerce system operator's transaction gains.
Consider $P_{ba} \!=\! 0.05$.
As $r_h$ drops from 200 to 100,
$G_s$ improves from 198.059 to 1142.670,
an improvement ratio of 4.77.
This improvement ratio also holds for
the average per seller transaction grains $G_e$.
It is interesting to observe that
when $P_{ba}$ is small,
both $G_s$ and $G_e$ are quite small.
In fact when $P_{ba} \!=\! 0.01$,
the $G_s$ is around 26.833
and $G_e$ is around 2.6833.
Namely, if buyers are less willing to
buy from ``average labeled'' sellers,
sellers (E-commerce operators) will have low long term profit gains
(average per seller transaction gains).

\begin{table}[htb]
\centering
\begin{tabular}{|c|c|c|c|c|c|}
\hline
$P_{ba}$ & 0.01   & 0.02   & 0.03   & 0.04 & 0.05 \\ \hline \hline
$G_s$ ($r_h \!=\! 200$)
  & 26.833  & 53.342  & 80.705 & 107.312 & 198.059 \\ \hline
$G_s$ ($r_h \!=\! 150$)
  & 26.980  & 53.594  & 80.812 & 369.951 & 1006.017 \\ \hline
$G_s$ ($r_h \!=\! 100$)
  & 26.941  & 54.433  & 760.511   & 1054.507 &  1142.670 \\ \hline \hline
$G_e$ ($r_h \!=\! 200$)
  & 2.6833  & 5.3342  & 8.0705 & 10.7312 & 19.8059 \\ \hline
$G_e$ ($r_h \!=\! 150$)
  & 2.6980  & 5.3594  & 8.0812 & 36.9951 & 100.6017 \\ \hline
$G_e$ ($r_h \!=\! 100$)
  & 2.6941  & 5.4433  & 76.0511 & 105.4507 &  114.2670 \\ \hline
\end{tabular}
\vspace{0.05 in}
\caption{Long term expected profit gains $G_s$
  and average per seller transaction gains $G_e$
  ($\lambda_1 = 20, \lambda_2 = 50,
  u =1, T=\alpha p = 0.1, \delta = 0.99,
  T_w = 180, P_{br} = 0.1, d=3$).}
\label{Tb:Reputation:Revenue}
\end{table}

\noindent
{\bf Summary: }
The reputation threshold $r_h$ and
$P_{ba}$
are critical to the ramp up time,
the new seller drop out probability,
the long term profit gains and
the average per seller transaction gains.
The baseline (or eBay-like) reputation system
presented in Section~\ref{section: model}
suffers from long ramp up time,
high seller drop out probability,
small long term profit gains and
small average per seller transaction gains.
Hence, it is important to ask
whether we can design a new mechanism
that an E-commerce system can use to
improve all the performance measures
$E[T_r]$, $P_d$, $G_s$ and $G_e$.
We next explore this interesting question.

\section{{\bf Insurance Mechanism}} \label{section: insurance}

In the previous section, we showed that the
baseline reputation system is not efficient.
Here, we present a new approach which we call the
``{\em insurance mechanism}'' to reduce both the
expected ramp up time ($E[T_r]$) and
new seller drop out probability ($P_d$), and to improve
the long term expected profit gains ($G_s$)
and the average per seller transaction gains ($G_e$).
We also quantify the impact of our insurance mechanism on
$E[T_r], P_d, G_s$ and $G_e$.
We show that our insurance mechanism
can reduce the ramp up time by around 87.2\%,
and guarantee new sellers ramp up before the deadline $T_w$
with a high probability (close to 1.0).
It also increases the long term profit gains
and average per seller transaction gains
by at least 95.3\%.

\subsection{{\bf Insurance Mechanism Design}}

The objective of our insurance mechanism
is to help new sellers ramp up quickly.
Reducing ramp up time brings the benefit
of reducing new seller drop out probability
and improving long term expected profit gains and
average per seller transaction gains
($G_s$ and $G_e$).
Our insurance mechanism consists of an {\em insurance protocol}
and a {\em transaction mechanism}.

We first describe the {\em insurance protocol}.
The E-commerce system operator provides an
insurance service to new sellers.
Each insurance has a price of $C_I \!>\! 0$,
a duration time of $T_d \!>\! 0$,
and a clearing time of $T_c \!>\!0$.
The insurance clearing time takes effect when
an insurance expires.
To buy an insurance,
a seller must provide the E-commerce operator
an insurance deposit of $D_I$.
Hence, the total payment by the new seller
to the E-commerce system operator is $C_I + D_I$.
We refer to this insurance as
the $(C_I, T_d, T_c, D_I)$-insurance.
Only new sellers can subscribe to this insurance.
If a seller subscribes an insurance,
the E-commerce system operator issues an
insurance certificate to him,
and this certificate is known to the public
(i.e., all buyers and sellers).
This certificate only takes effect
within the insurance duration time $T_d$.
The E-commerce system operator treats a seller
with an insurance certificate as trustworthy.
To guarantee that such sellers
will advertise their product quality honestly,
the E-commerce system operator
requires such sellers obey
the following {\em transaction mechanism}.

We now describe the {\em transaction mechanism}.
Only sellers with an insurance certificate must
obey this transaction mechanism.
Let us focus on a seller with
an insurance certificate.
When ordering a product from this seller,
a buyer sends his payment $p$ to
the E-commerce system operator.
After receiving the product,
if this buyer express a positive feedback,
then this transaction completes,
i.e., the E-commerce operator
forwards the payment $(1 - \alpha) p$ to the seller and
charges a transaction fee of $\alpha p$.
This transaction also completes if this
buyer expresses a neutral feedback.
A neutral feedback means that
a seller slightly overstated his product quality,
i.e. $Q_i \!<\! Q_a \!<\! Q_i +\gamma$.
To avoid such overstating,
the E-commerce company revokes a seller's
insurance certificate
once the fraction of positive feedbacks
falls below the consistency factor ($\theta$),
i.e., $n^+ / (n^+ + n^0 + n^-) < \theta$.
A negative feedback results in the
transaction being revoked.
More concretely, the E-commerce operator gives
the payment $p$ back to the buyer
and does not charge
any transaction fee from the seller
(provided that it is within
the duration time $T_d$,
or the clearing time $T_c$).
The buyer needs to ship the product
back to the seller but the buyer does not need to pay
for the shipment cost $C_S$,
for it will be deducted from
a seller's insurance deposit $D_I$.
If the insurance deposit is not
enough to cover $C_S$,
the E-commerce operator makes a
supplemental payment.
To avoid this undesirable outcome,
the E-commerce company revokes a seller's
insurance certificate,
once a seller's deposit reaches a threshold
$\WHD_I \!<\! D_I$.
The insurance clearing time
takes effect when an insurance is invoked.
At the end of the clearing time,
the E-commerce company returns the
remaining deposit of $D_I$
(if it is not deducted to zero) back to the seller.

\noindent
{\bf Remark.}
Sellers may collude with buyers
to inflate their reputation by fake transactions \cite{Hoffman2009}.
One way to avoid such collusion is
by increasing the transaction fee
such as~\cite{Bhattacharjee2005}.
The shipment cost may exceed $D_I$
due to a large number of products to be returned.
This can be avoided with high probability
by setting a large $\WHD_I$
(Theorem~\ref{Mechanism:thm:InsurancePrice}).
We also derive the minimum clearing time ($T_c$)
to guarantee
that a seller with an insurance certificate
needs to obey the transaction mechanism
(Theorem~\ref{Mechanism:thm:InsurancePrice}).

\subsection{{\bf Analyzing the Insurance Mechanism}}

We first show that buyers treat
a seller having an insurance certificate
as trustworthy.
Through this, we derive the transaction rate
that a seller with an insurance certificate can attract.
We then derive the improved $E[T_r], P_d, G_s$ and $G_e$.

Buyers treat sellers having an insurance certificate as trustworthy.
This is an important property of our insurance mechanism
because it influences the probability that a buyer
adopts a product from a seller.
Suppose in time slot $\tau$,
a seller has an insurance certificate.
If this seller advertise honestly $Q_a \!=\! Q_i$,
then the buyer who buys a product from this seller
will be satisfied (express positive feedback rating).
In this case, the payment from the buyer will
be forwarded to the seller.
Hence this seller earns a profit of $u$.
If this seller overstates his product quality
beyond the lenient factor ($\gamma$),
i.e., $Q_a \!>\! Q_i + \gamma$.
Then according to our insurance mechanism,
the payment by the buyer will be returned back to the buyer.
The seller needs to pay a shipment cost of $C_S$ to
ship back the product and $C_S$ will be deducted
from his insurance deposit $D_I$.
Hence, if a seller overstates the product quality
beyond the lenient factor,
he will lose a total shipment cost of at least
$\min\{\WHD_I, C_S N(\tau)\}$
in time slot $\tau$,
where $N(\tau)$ denotes the number of product selling.
A seller with an insurance certificate must obey
the same consistency factor ($\theta$)
as reputable sellers in being honest,
i.e., $n^+ / (n^+ + n^0 + n^-) \!\geq\! \theta$,
because if not his insurance certificate will
be revoked by the E-commerce operator.
Given these properties, buyers trust a seller
with an insurance certificate.
Recall that the E-commerce operator also trusts a seller
with an insurance certificate.
Therefore, an insured  seller can attract transactions
with an arrival rate being equivalent to those reputable sellers.
Let $\lambda_T^I$ denote the transaction's arrival rate
to a seller with an insurance certificate.
We have
\begin{equation}
\lambda_T^I \!=\!\lambda_2 P_{br}.
\label{Mechanism:Eq:TransactionRate}
\end{equation}

We now quantify the impact of our insurance mechanism
on the four performance measures.
Let $T_r^I, P_d^I, G_s^I,G_e^I$ denote the
ramp up time, the new seller drop out probability,
the long term profit gains and
the average per seller transaction gains
respectively,
when a new seller subscribes our insurance.

\begin{theorem}
Suppose a new seller subscribes to
our proposed insurance mechanism.
We express the expected ramp up time
and new seller drop out probability as
\begin{align*}
& E[T_r]
  = d \sum_{\tau=1}^\infty
  \left( 1 - \sum_{k=r_h}^{\infty}
  e^{- \sum_{\ell=0}^{\tau-2} \WTlmda_T(\ell)}
  \frac{(\sum_{\ell=0}^{\tau-2} \WTlmda_T(\ell))^k}{k!}
  \right), \\
& P_d
  =
  \sum_{k=0}^{r_h - 1}
  e^{- \sum_{\ell=0}^{T_w/d-1} \WTlmda_T(\ell)}
  \frac{(\sum_{\ell=0}^{T_w/d -1} \WTlmda_T(\ell))^k}{k!},
\end{align*}
where $\WTlmda_T(\ell) \!=\! \lambda_2 P_{br} d$
for all $\ell \!=\! 0,1,\ldots, \lfloor T_d/d \rfloor -1$,
and $\WTlmda_T(\lfloor T_d/d \rfloor) \!=\!
\lambda_2 P_{br} (T_d - d \lfloor T_d/d \rfloor) +
\lambda_1 P_{ba} (d \lfloor T_d/d \rfloor + d - T_d)$,
and $\WTlmda_T(\ell) \!=\! \lambda_1 P_{ba} d$
for all $\ell \!=\! \lfloor T_d/d \rfloor +1, \ldots, \infty$.
The long term expected profit gains
for an insured seller is:
\begin{align*}
& G_s^I
  =
  \sum_{k = 0}^{r_h -1}
  \frac{(\lambda_2 P_{br} \min\{T_d, T_w\})^{k}}
    {k!}
  e^{- \lambda_2 P_{br} \min\{T_d, T_w\}}  \\
&
  \int \ldots \int\limits_{
  \hspace{-0.35 in}
  0< t_1 < \ldots < t_k < \min\{T_d, T_w\} }
  \!\!\!\!
  \frac{k! dt_1 \ldots d t_k}{(\min\{T_d, T_w\})^{k}}
  \bigg(\! \MBI_{\{T_d \geq T_w\}}  \!\!
  \sum_{j=1}^k u \delta^{\lceil t_j / d \rceil}\\
&
  +
  \MBI_{\{T_d < T_w\}}
  \sum_{i=0}^{r_h - 1 - k}
  \frac{(\lambda_1 P_{ba} (T_w - T_d))^i}{i!}
  e^{- \lambda_1 P_{ba} (T_w - T_d)}
  \\
&
  \int \ldots \int\limits_{
  \hspace{-0.35 in}
  T_d< t_{k+1} < \ldots < t_{k+i} < T_w}
  \!\!\!\!
  \frac{i!}{(T_w - T_d)^i}
  d t_{k+1} \ldots t_{k+i}
  \sum_{j=1}^{k+i} u \delta^{\lceil t_j / d \rceil}
  \bigg)
  \\
&
  +
  \bigg(
  \MBI_{\{T_d \geq T_w\}}
  \int_0^{T_w}
  \lambda_2 P_{br}
  e^{- \lambda_2 P_{br} t_{r_h}}
  \frac{(\lambda_2 P_{br} t_{r_h})^{r_h - 1}}
    {(r_h - 1)!}
    dt_{r_h}  \\
&
  \int \ldots \int\limits_{
  \hspace{-0.33 in}
  0< t_1 < \ldots < t_{r_h -1} < t_{r_h}}
  \frac{(r_h - 1)!}{(t_{r_h})^{r_h -1}}
  dt_1 \ldots dt_{r_h-1}
  +
  \MBI_{\{T_d < T_w\}}
  \sum_{k=0}^{r_h}
  \\
& \frac{(\lambda_2 P_{br} T_d)^k}{k!}
  e^{-\lambda_2 P_{br} T_d}
  \int \ldots \int\limits_{
  \hspace{-0.33 in}
  0< t_1 < \ldots < t_{k} < T_d}
  \!\!\!
  \frac{k!}{T_d^k}
  dt_1 \ldots dt_k
  \\
&   \int_{T_d}^{T_w}
  \lambda_1 P_{ba}
  e^{-\lambda_1 P_{ba} (t_{r_n} - T_d)}
  \frac{(\lambda_1 P_{ba} (t_{r_n} - T_d))^{r_h - k -1}}
    {(r_h - k -1)!} d t_{r_h}
  \\
&
  \int \ldots \int\limits_{
  \hspace{-0.33 in}
  T_d< t_{k+1} < \ldots < t_{r_h -1} < t_{r_h}}
  \!\!\!\!\!\!\!\!
  \frac{(r_h - k - 1)! dt_{k+1} \ldots dt_{r_h -1}}
    {(t_{r_h} - T_d)^{r_h - k - 1}}
  \!\!
  \bigg)
  u \bigg( \lambda_2 d P_{br}\\
&  \frac{\delta^{\lceil t_{r_h}/d \rceil + 1} }{1 - \delta}
  +
  \sum_{j=1}^{r_h} \delta^{\lceil t_j/d \rceil}
  +
  \delta^{\lceil t_{r_h}/d \rceil} \lambda_2 P_{br}
  (d \lceil t_{r_h} / d \rceil - t_{r_h})\!\!
  \bigg)
\end{align*}
Furthermore,
$G_e^I = \frac{\alpha p}{u} G_s^I$.
\label{Mechanism:thm:ImpactMec}
\end{theorem}

\noindent
{\bf Proof:}
Please refer to the appendix for derivation.
\done

\noindent
{\bf Remark:}
Theorem~\ref{Mechanism:thm:ImpactMec} quantifies
the impact of our insurance mechanism on the
four important performance measures.
Before we talk more about how to select the
insurance price $C_I$ and deposit $D_I$,
let us illustrate the effectiveness of our insurance
mechanism using some numerical examples.

Table~\ref{Tb:Mechanism:Result} presents numerical
examples on $E[T_r^I], P_d^I, G_s^I$ and $G_e^I$.
We use the following setting:
$\lambda_1 \!=\! 20, \lambda_2 \!=\! 50,
u \!=\! 1, T_w \!=\! 180, d \!=\! 3,
P_{br} \!=\! 0.1, P_{ba} \!=\! 0.03,
C_I \!=\! 100, D_I \!=\! 100,
\WHD_I \!=\! 50, C_S \!=\! 0.5,
T \!=\! 0.1, T_d \!=\! 100, T_c \!=\! 3,
\delta \!=\! 0.99$.
We also presents numerical examples on
$E[T_r], P_d, G_s$ and $G_e$ for comparison studies.
When $r_h \!=\! 100$,
we have $E[T_r] \!=\! 168.1$ and $E[T_r^I] \!=\! 21.5$.
In other words, our insurance mechanism reduces the
expected ramp up time from 168.1 days
to only 21.5 days, or over 87.2\% reduction.
It is interesting to observe that our incentive mechanism
reduces the new seller drop out probability
from $P_d = 0.20819$ to $P_d^I = 0$.
Namely, our insurance mechanism can guarantee that
new sellers ramps up before the deadline line $T_w$
with a high probability (very close to 1.0).
In addition, our insurance mechanism improves
long term expected profit gains from $G_s = 760.51$
to $G_s^I = 1485.04$, a 95.3\% improvement.
This improvement ratio also holds for
average per seller transaction gains.
As $r_h$
increases from 100 to 200,
the improvement on the $E[T_r], P_d, G_s, G_e$,
becomes more significant.
We next state the appropriate values
for $C_I, D_I, \WHD_I$ and $T_c$
in the following theorem.

\begin{table}[htb]
\centering
\begin{tabular}{|c||c|c|c|}
\hline
    & $r_h \!=\! 100$   & $r_h \!=\! 150$ & $r_h \!=\! 200$ \\ \hline \hline
$(E[T_r], E[T_r^I])$
  & (168.1, 21.5) & (251.6, 31.5) & (334.9, 41.5) \\ \hline
$(P_d, P_d^I)$
  & (0.20819, 0)  & (0.99992, 0) & (1.0, 0) \\ \hline
$(G_s, G_s^I)$
  & (760.51, 1485.04)  & (80.81, 1485.03) & (80.71, 1485.01) \\ \hline
$(G_e, G_e^I)$
  & (76.051, 148.504)  & (8.081, 148.503) & (8.071, 148.501) \\ \hline
\end{tabular}
\vspace{0.05 in}
\caption{Impact of our insurance on $E[T_r]$,
  $P_d, G_s$ and $G_e$.}
\label{Tb:Mechanism:Result}
\end{table}


\begin{theorem}
An upper bound for the insurance price $C_I$ is
$
 C_I < G_s^I - G_s.
$
If $D_I$ and $\WHD_I$ satisfies
\[
D_I > \WHD_I \geq C_S
\max\{\ln \epsilon^{-1} - \lambda_2 P_{br} T_d,
e^2 \lambda_2 P_{br} T_d\},
\]
then $\Pr[\mbox{shipment cost exceeds $D_I$}] \leq \epsilon$.
If $T_c \!\geq\! d$,
then all products sold by a seller with an insurance certificate
can be guaranteed to obey the insurance mechanism.
\label{Mechanism:thm:InsurancePrice}
\end{theorem}

\noindent
{\bf Proof:}
Please refer to the appendix for derivation.
\done

\noindent
{\bf Remark:}
The insurance price should be lower than $G_s^I - G_s$.
The clearing time should be larger or equal to $d$.
To guarantee that
the insurance deposit covers the shipment cost for
returning products with high probability,
$D_I$ and $\WHD_I$ need to be no less than
$C_S
\max\{\ln \epsilon^{-1} - \lambda_2 P_{br} T_d,
e^2 \lambda_2 P_{br} T_d\}$.

\section{{\bf Related Work}}
\label{section: relatedwork}

Research on reputation systems~\cite{Resnick2000}
for internet services has been quite active.
Many aspects of reputation systems have been studied, i.e.,
reputation metric formulation and calculation
\cite{Houser2006,Kamvar2003,Resnick2009},
attacks and defense techniques for reputation systems
\cite{Cheng2005,Hoffman2009,Yu2006,Viswanath2012},
and effectiveness of reputation systems~\cite{Dellarocas2000}.
A survey can be found in~\cite{Josang2007}.

Theoretical aspects of reputation system have been studied extensively.
First, many works studied reputation metric formulation and calculation.
Two most representative reputation calculating models are
the eBay-like reputation model~\cite{Houser2006}
and the transitive trust based model~\cite{Cheng2005}.
The eBay-like reputation system is a typical
example of reputation model which computes
the reputation score by summarizing explicit
human feedbacks (or ratings)~\cite{Buchegger2004,Houser2006,Singh2003,Zhou2007}.
The transitive trust based
model~\cite{Cheng2005,Delaviz2012,Resnick2009,Kamvar2003,Yu2006}
assumes that if user $A$ trusts user $B$ and
user $B$ trusts user $C$, then user $A$ trusts user $C$.
More precisely, each user is represented by a node in a graph,
and the weighted directed link from $A$ to $B$ quantifies
the degree that user $A$ trusts user $B$.
For this model, many algorithms were developed to
compute an overall reputation score for each user~\cite{Cheng2005,Delaviz2012,Resnick2009,Kamvar2003,Yu2006}.
These works provided theoretical foundations
for reputation computing.
Second, many works explored
attack and defense techniques for reputation systems.
One type of potential attacks
is that users may not give honest feedbacks.
Peer-prediction method based mechanisms were proposed
to elicit honest feedbacks~\cite{Jurca2006,Jurca2007,Miller2005}.
Another type of potential attacks is reputation inflation,
or self-promotion.
Many works have been done to address this issue
\cite{Cheng2005,Hoffman2009,Yu2006,Viswanath2012}.
A survey on attack and defense techniques
for reputation systems can be found in~\cite{Hoffman2009}.
The main difference between our work and theirs is that
we propose a new mechanism to improve eBay system.

The most closely related works are
\cite{Bhattacharjee2005,Dellarocas2000,Dellarocas2001,Khopkar2005},
which studied the eBay reputation mechanism.
Authors in~\cite{Bhattacharjee2005} derived the minimum
transaction fee to avoid ballot stuffing
(i.e., fake positive feedbacks).
Authors in~\cite{Dellarocas2000} proposed an algorithm
based on buyer friendship relationship to filter out
unfair ratings.
In~\cite{Dellarocas2000}, authors explored
the impact of buyers biases'
(i.e., leniency or criticality) in express feedback ratings
on sellers in advertising product quality.
The impact of negative feedbacks on buyers
in expressing feedback ratings was studied in~\cite{Khopkar2005}.
The difference between our work and theirs is that
we propose a new mechanism to improve eBay system.

\section{{\bf Conclusion}} \label{section: conclusion}

This paper presents an Insurance mechanism to improve
eBay-like reputation mechanisms.
We proposed four performance measures to
analyze eBay reputation system:
(1) new seller ramp up time,
(2) new seller drop out probability,
(3) long term profit gains for sellers and
(4) average per seller transaction gains for an E-commerce operator.
We developed a stochastic model to identify key factors which
influence the above four measures.
We proposed an insurance mechanism to
improve the above four measures.
We show that our insurance mechanism
can reduce the ramp up time by around 87.2\%,
and guarantee new sellers ramp up before the deadline $T_w$
with a high probability (close to 1.0).
It also increases the long term profit gains
and average per seller transaction gains
by at least 95.3\%.

\bibliographystyle{abbrv}
\bibliography{reference}

\section*{{\bf Appendix}}

We first state a lemma which will be used in our proof.
\begin{lemma}
[\cite{Chen2002}]
Let $\{N'(t), t \geq 0\}$ denote a Poisson process with
a rate parameter $\lambda$.
Let $t_k$ denote the arrival time of $k$-th event.
Let $f(t_1,\ldots,t_n | N'(t') = n)$ denote the
conditional probability density function of $t_1,\ldots, t_n$
given $N'(t') = n$.
Then we have
$f(t_1,\ldots,t_n | N'(t') = n) = n! / (t')^{n}$,
where $0<t_1 <\ldots<t_n<t'$.
\label{Proof:lem:probDens}
\end{lemma}

\noindent
{\bf Proof of Theorem~\ref{Reputation:Thm:Rampuptime}:}
Note that each new seller advertise product quality
honestly.
In this scenario, each transaction results in
one positive feedback.
Recall our reputation updating rule
specified in Equation~(\ref{Model:Eq:reputationUpdating}),
we have that the reputation score at time slot $\tau$
equals the number of transactions arriving within
time slot 0 to time slot $\tau -1$.
Recall the definition of $T_r$ in
Equation~(\ref{Problem:Eq:RampTime}),
we have that $T_r / d \in \mathbb{N}$.
With these observations and
by some basic probability arguments,
we have
\begin{align*}
& E[T_r]
  = \sum_{\tau=1}^\infty \tau d \Pr[T_r = \tau d]
  = d \sum_{\tau=1}^\infty \tau \Pr[T_r/d = \tau ]  \\
&
  = d \sum_{\tau=1}^\infty \Pr[T_r/d \geq \tau ]
  = d \sum_{\tau=1}^\infty (1 - \Pr[T_r/d \leq (\tau -1)])\\
&
  = d \sum_{\tau=1}^\infty (1 - \Pr[r(\tau -1) \geq r_h])
  = d \sum_{\tau=1}^\infty \!\!
  \left(\!1 \!- \!\!
  \sum_{\ell=0}^{\tau-2} N(\ell) \!\geq\! r_h \!\right).
\end{align*}
Note that $\sum_{\ell=0}^{\tau-2} N(\ell)$ is
a random variable which follows a Poisson distribution
with parameter $\lambda_1 P_{ba} (\tau - 1) d$.
We have
\begin{align*}
E[T_r]
& = d \sum_{\tau=1}^\infty (1 - \sum_{k=r_h}^\infty
  e^{- \lambda_1 P_{ba} (\tau-1) d}
  \frac{(\lambda_1 P_{ba} (\tau -1 ) d )^k}{k!}).
\end{align*}
Evaluating the first order derivative on $E[T_r]$
with respect to $r_h$ and $\lambda_1 P_{ba}$
respectively, one can easily obtain the monotonous
property of $E[T_r]$.
\done

\noindent
{\bf Proof of Theorem~\ref{Reputation:Thm:BankruptProb}:}
Applying similar derivation as
Theorem~\ref{Reputation:Thm:Rampuptime},
we have that the reputation score at time slot $\tau$
equals the number of transactions arriving within
time slot 0 to $\tau -1$.
Note that $T_w / d \in \mathbb{N}$.
Recall the definition in
Equation~(\ref{Problem:Eq:RampTime})
we have that $T_r > T_w$ if and only if
$r(T_w/d ) < r_h$.
Using some basic probability arguments,
we have
\[
P_d
\!=\! \Pr[r(T_w / d ) < r_h]
\!=\! \Pr\left[\sum\nolimits_{\tau =0}^{T_w / d -1}
\!N(\tau) < r_h\right] .
\]
Note that $\sum_{\tau =0}^{T_w / d  -1} N(\tau)$
is a random variable which follows a Poisson distribution
with parameter $\lambda_1 P_{ba} T_w $.
We have
\begin{align*}
P_d
& \!=\!\! \sum_{k=0}^{r_h-1}
  \!\Pr \!\left[\sum_{\tau =0}^{T_w / d -1} \!\!\!\!
  N(\tau) = k\right]
 \!=\! \sum_{k=0}^{r_h-1}\!\!
  e^{- \lambda_1 P_{ba} T_w }
  \frac{(\lambda_1 P_{ba} T_w )^k}{k!} .
\end{align*}
Evaluating the first order derivative on $P_d$
with respect to $r_h$, $T_w$ and $\lambda_1 P_{ba}$
respectively, one can easily obtain the monotonous
property of $P_d$.
\done

\noindent
{\bf Proof of Theorem~\ref{Reputation:Thm:Revenue}:}
Let $\WHG_s \!=\! \sum_{\tau = 0}^\infty \delta^{\tau} u N (\tau)$.
Then $\WHG_s$ is a random variable and
$G_s \!=\! E[\WHG_s]$.
Based on whether a seller ramps up or not,
we divide $E[\WHG_s]$ into two parts, i.e.,
\begin{align*}
E[\WHG_s] =
& \Pr[T_r > T_w] E[\WHG_s | T_r > T_w]  \\
& +
  \Pr[T_r \leq T_w] E[\WHG_s | T_r \leq T_w]
\end{align*}
We next derive the above two terms individually.

We first derive $\Pr[T_r > T_w] E[\WHG_s | T_r > T_w]$.
Note that each new seller advertise product quality honestly.
Hence each transaction earns one positive feedback.
Recall that $T_w / d \in \mathbb{N}$.
Note that at time slot $T_w/d $ a seller drops out,
i.e., there will be no transactions from time slot $T_w/d$.
Let $K$ denote the number of transactions
arriving within time slot 0 to $T_w/d - 1$.
Then $K$ satisfies $0 \leq K \leq r_h -1$
since we are given $T_r > T_w$.
Note that $K = k$, where $k \leq r_h - 1$ implies that
$T_r > T_w$ and $K$ is of a value larger than $r_h -1$
implies that $T_r \leq T_w$.
Then we have
\begin{align*}
& \Pr[T_r > T_w] E[\WHG_s | T_r > T_w] \\
& =
  \sum\nolimits_{k = 0}^{r_h -1}
  \Pr\left[K = k, T_r > T_w \right]
  E[\WHG_s | T_r > T_w, K=k]  \\
& = \sum\nolimits_{k = 0}^{r_h -1}
  \Pr\left[K = k \right]
  E[\WHG_s |  K=k].
\end{align*}
Observe that $K \!=\! \sum_{\tau=0}^{T_w/d -1} N(\tau)$
is a random variable which follows
a Poisson distribution with parameter
$\lambda_1 P_{ba} T_w$.  We then have
\begin{align*}
& \Pr[T_r > T_w] E[\WHG_s | T_r > T_w] \\
& \hspace{0.17 in}
  = \sum_{k = 0}^{r_h -1}
  e^{- \lambda_1 P_{ba} T_w}
  \frac{(\lambda_1 P_{ba} T_w)^k}{k!}
  E[\WHG_s | K=k].
\end{align*}
We next derive $E[\WHG_s | K=k]$.
Let $t_1, \ldots, t_K$
denote the arrival time of transaction $1,\ldots,K$.
Then $t_1, \ldots, t_K$ satisfy
$0 < t_1 < \ldots < t_K < T_w$.
Let $f(t_1, \ldots,t_k | K=k)$ denote the probability
density function of $t_1, \ldots,t_k$
given that $K=k$.
By applying Lemma~\ref{Proof:lem:probDens},
we obtain that
$f(t_1,\ldots,t_k | K=k) = k! / (T_w)^{k}$.
Then we have
\begin{align*}
& E[\WHG_s | K=k]
  \\
& =
  \int \ldots \int\limits_{
  \hspace{-0.35 in}
  0 < t_1 < \ldots < t_k < T_w}
  f(t_1,\ldots,t_k | K=k)
  E[\WHG_s | t_1, \ldots, t_k] \\
& \hspace{0.17 in}
  dt_1 \ldots dt_k \\
& =
  \int \ldots \int\limits_{
  \hspace{-0.35 in}
  0 < t_1 < \ldots < t_k < T_w }
  \frac{k!}{T_w^{k}}
  \sum_{j=1}^k u \delta^{\lceil t_j / d \rceil}
  dt_1 \ldots dt_k,
\end{align*}
where the first step follows that given
$t_1, \ldots, t_k$ is equivalent to given $K=k$.
The second step follows that
the payment of transaction $1\leq j \leq k$
is forwarded to the seller in time slot
$\lceil t_j / d \rceil$ because of the shipment delay.
Namely, in computing of the long term profit gains,
the $j$-th transaction results in a discounted
profit gain of $u \delta^{\lceil t_j / d \rceil}$.
Hence
$E[\WHG_s | t_1, \ldots, t_k] =
\sum_{j=1}^k u \delta^{\lceil t_j / d \rceil}$.
Combining them all, we have
\begin{align*}
& \Pr[T_r > T_w] E[\WHG_s | T_r > T_w]
  =\!\! \sum_{k = 0}^{r_h -1}
  e^{- \lambda_1 P_{ba} T_w}
  \frac{(\lambda_1 P_{ba} T_w)^k}{k!}\\
& \hspace{0.17 in}
  \int \ldots \int\limits_{
  \hspace{-0.35 in}
  0 < t_1 < \ldots < t_k < T_w }
  \frac{k!}{T_w^{k}}
  \sum_{j=1}^k u \delta^{\lceil t_j / d \rceil}
  dt_1 \ldots dt_k.
\end{align*}

We now derive $\Pr[T_r \leq T_w] E[\WHG_s | T_r \leq T_w]$.
Let $t_{r_h}$ denote the arrival time of the $r_h$-th transaction.
Based on our reputation updating rule specified
in Equation~(\ref{Model:Eq:reputationUpdating}),
we obtain that $t_{r_h} < T_w$ implies
$T_r \leq T_w$.
Note that $t_{r_h}$ is a random variable.
Let $f(t_{r_h})$ denote the probability
density function of $t_{r_h}$.
Observe that
$\Pr[t_{r_h} \leq t]
= \sum_{k=r_h}^\infty e^{- \lambda_1 P_{ba} t}
(\lambda_1 P_{ba} t)^k / k!.
$
Performing the first order derivative on this term, we have
$f(t_{r_h}) \!=\! \lambda_1 P_{ba}
e^{- \lambda_1 P_{ba} t_{r_h}}
(\lambda_1 P_{ba} t_{r_h})^{r_h - 1} / (r_h -1)!.$
Then it follows that
\begin{align*}
& \Pr[T_r \!\leq\! T_w] E[\WHG_s | T_r \!\leq\! T_w]
\!=\!\! \int_{0}^{T_w } \!\!\! f(t_{r_h})
  E[\WHG_s |  t_{r_h}] d t_{r_h} \\
& = \int_{0}^{T_w }
  \!\!\!\!\!\! \lambda_1 P_{ba}
  e^{- \lambda_1 P_{ba} t_{r_h}}
  \frac{(\lambda_1 P_{ba} t_{r_h})^{r_h - 1} }
    { (r_h -1)! }
  E[\WHG_s | t_{r_h}] d t_{r_h}.
\end{align*}
We next derive $E[\WHG_s | t_{r_h}]$.
Let $t_1,\ldots, t_{r_h -1}$ denote the arrival time
of the $1$-st,$\ldots, (r_h-1)$-th transaction.
Let $f(t_1,\ldots, t_{r_h -1} | t_{r_h})$
denote the probability density function of
$t_1,\ldots, t_{r_h -1}$ given $t_{r_h}$.
Then applying Lemma~\ref{Proof:lem:probDens},
we obtain that
$f(t_1,\ldots, t_{r_h -1} | t_{r_h})
= (r_h -1)! / (t_{r_h})^{r_h - 1}$.
Then it follows that
\begin{align*}
&E[\WHG_s | t_{r_h}]
= \int \ldots \int\limits_{
  \hspace{-0.35 in}
  0 < t_1 < \ldots < t_{r_h-1} < t_{r_h} }
  \!\!\!\!\!\!\!\!\!
  f(t_1,\ldots, t_{r_h -1} | t_{r_h}) \\
& \hspace{0.17 in}
  E[\WHG_s | t_1,\ldots, t_{r_h -1}, t_{r_h}]
  dt_1 \ldots dt_{r_h -1}  \\
&
  =
  \int \ldots \int\limits_{
  \hspace{-0.35 in}
  0 < t_1 < \ldots < t_{r_h-1} < t_{r_h} }
  \frac{(r_h -1)!}{(t_{r_h})^{r_h - 1} }
  u \Big( \sum_{j =1}^{r_h} \delta^{\lceil t_j / d \rceil} +
  \delta^{\lceil t_{r_h} / d \rceil} \\
& \hspace{0.17 in}
  \lambda_1 P_{ba} (d \lceil t_{r_h} / d \rceil - t_{r_h})
  \! +\!  \lambda_2 P_{br} d
  \frac{\delta^{\lceil t_{r_h}/d \rceil + 1}}{1 - \delta}
  \Big)
  dt_1 \ldots dt_{r_h -1}.
\end{align*}
We elaborate more on computing
$E[\WHG_s | t_1,\ldots, t_{r_h -1}, t_{r_h}]$.
The transactions arriving within time
$0$ to time $t_{r_h}$ contribute
$\sum_{j =1}^{r_h} u \delta^{\lceil t_j / d \rceil}$
to the long term profit gains.
Consider the transactions arriving within time
$t_{r_h}$ to $d \lceil t_{r_h} / d \rceil$.
This time interval belongs to time slot
$\lfloor t_{r_h} / d \rfloor$.
Note that in this time slot the reputation
score is lower than $r_h$.
This means that the number of transactions arriving in this
time interval follows a Poisson distribution
with parameter
$\lambda_1 P_{ba} (d \lceil t_{r_h} / d \rceil - t_{r_h})$.
In expectation, transactions arriving in this time
interval contribute
$\delta^{\lceil t_{r_h} / d \rceil}
\lambda_1 P_{ba} (d \lceil t_{r_h} / d \rceil - t_{r_h})u$.
Consider transactions arriving from
time slots $\lceil t_{r_h} / d \rceil$ to $\infty$.
In these time slots, a seller's reputation satisfies the condition to
be labeled as reputable.
The number of transactions arriving in each of
these time slots follows a Poisson distribution with parameter
$\lambda_2 P_{br} d$.
These transactions, in expectation, contribute
$u \lambda_2 P_{br} d
\frac{\delta^{\lceil t_{r_h}/d \rceil + 1}}{1 - \delta}$
in total.
Summing these terms together we obtain
$E[\WHG_s | t_1,\ldots, t_{r_h -1}, t_{r_h}]$.
We have
\begin{align*}
& \Pr[T_r \!\leq\! T_w] E[\WHG_s | T_r \!\leq\! T_w] \\
& = \int_{0}^{T_w}
  \lambda_1 P_{ba}
  e^{- \lambda_1 P_{ba} t_{r_h}}
  \frac{(\lambda_1 P_{ba} t_{r_h})^{r_h - 1} }
    { (r_h -1)! }
  d t_{r_h} \\
& \hspace{0.17 in}
  \int \ldots \int\limits_{
  \hspace{-0.35 in}
  0 < t_1 < \ldots < t_{r_h-1} < t_{r_h} }
  \frac{(r_h -1)!}{(t_{r_h})^{r_h - 1} }
  u \Big( \sum_{j =1}^{r_h} \delta^{\lceil t_j / d \rceil} +
  \frac{\delta^{\lceil t_{r_h}/d \rceil + 1}}{1 - \delta}
   \\
& \hspace{0.17 in}
  \lambda_2 P_{br} d
  +
  \delta^{\lceil t_{r_h} / d \rceil}
  \lambda_1 P_{ba} (d \lceil t_{r_h} / d \rceil - t_{r_h})
  \Big)
  dt_1 \ldots dt_{r_h -1}
\end{align*}
Combing them all, we prove this theorem.
\done

\noindent
{\bf Proof of Theorem~\ref{Mechanism:thm:ImpactMec}:}
We first derive $E[T_r^I]$ and $P_d^I$.
Note that sellers advertise honestly.
This means that all transactions result in
positive feedbacks.
This implies that the insurance certificate
expires at the end of the duration time.
Note that $N(\ell)$, the number of transactions at time slot
$\ell \!=\!0,1,\ldots,\infty$ before a seller ramps up,
follows a Poisson distribution,
and we denote its parameter by $\WTlmda_T(\ell)$.
Applying Equation~(\ref{Mechanism:Eq:TransactionRate}),
we have
$\WTlmda_T(\ell) \!=\! \lambda_2 P_{br} d$
for all $\ell \!=\! 0,1,\ldots, \lfloor T_d/d \rfloor -1$,
and $\WTlmda_T(\lfloor T_d/d \rfloor) \!=\!
\lambda_2 P_{br} (T_d - d \lfloor T_d/d \rfloor) +
\lambda_1 P_{ba} (d \lfloor T_d/d \rfloor + d - T_d)$,
and $\WTlmda_T(\ell) \!=\! \lambda_1 P_{ba} d$
for all $\ell \!=\! \lfloor T_d/d \rfloor +1, \ldots, \infty$.
Then with a similar derivation as
Theorem~\ref{Reputation:Thm:Rampuptime}
we obtain the expected ramp up time $E[T_r^I]$.
Furthermore, with a similar derivation as
Theorem~\ref{Reputation:Thm:BankruptProb}
we obtain $P_d^I$.

Let us now derive long term profit gains ($G_s$)
and average per seller transaction gains ($G_e$).
We derive $G_s$ first.
Let $\WHG_s^I \!=\! \sum_{\tau = 0}^\infty \delta^{\tau} u N^I (\tau)$,
where $N^I (\tau)$ denotes the number of transactions
arriving in time slot $\tau$ when a seller
subscribes to an insurance.
Then $\WHG_s^I$ is a random variable and
$G_s^I \!=\! E[\WHG_s^I]$.
With a similar derivation as
Theorem~\ref{Reputation:Thm:Revenue},
we have
\begin{align*}
E[\WHG_s^I] =
& \Pr[T_r^I > T_w] E[\WHG_s^I | T_r^I > T_w]  \\
& +
  \Pr[T_r^I \leq T_w] E[\WHG_s^I | T_r^I \leq T_w]
\end{align*}
Let $\WDlmda_T (t)$ denote the transaction's
arrival rate at time $t [0, \infty)$.
Applying Equation~(\ref{Mechanism:Eq:TransactionRate}),
we have two cases:
(1) a seller ramps up ($T_r^I \leq T_w$),
then we have
$\WDlmda_T (t) \!=\! \lambda_2 P_{br}$
  for all $t \in [0, \min\{T_w, T_d\}]$,
$\WDlmda_T (t) \!=\! \lambda_1 P_{ba}$
  for all $t \in [T_d, T_r^I]$,
and $\WDlmda_T (t) \!=\! \lambda_2 P_{br}$
  for all $t \in [T_r^I, \infty]$;
(2) a seller drops out ($T_r^I > T_w$),
then we have
$\WDlmda_T (t) \!=\! \lambda_2 P_{br}$
  for all $t \in [0, \min\{T_w,T_d\}]$,
$\WDlmda_T (t) \!=\! \lambda_1 P_{ba}$
  for all $t \in [T_d, T_w]$,
and $\WDlmda_T (t) \!=\! 0$
  for all $t \in [T_w, \infty]$.
With these observations and using a similar derivation as
Theorem~\ref{Reputation:Thm:Revenue},
one can easily obtain analytical expressions
for the term
$\Pr[T_r^I > T_w] E[\WHG_s^I | T_r^I > T_w]$
and the term
$\Pr[T_r^I \leq T_w] E[\WHG_s^I | T_r^I \leq T_w]$
respectively.
Combing them all we complete this theorem.
\done

\noindent
{\bf Proof of Theorem~\ref{Mechanism:thm:InsurancePrice}:}
We want to derive the reasonable price
that an E-Commerce operator can charge for the insurance.
The marginal long term profit gain of an insured seller is
$G^I_s - C_I$.
Note that the marginal long term profit gain
without insurance is $G_s$.
Thus sellers has the incentive to buy an insurance
if the marginal profit gain corresponds to
buying an insurance is
larger than the marginal profit gain without
insurance, i.e., $G^I_s - C_I \!>\! G_s$,
which yields $C_I \!<\! G^I_s - G_s$.

Note that sellers advertise honestly.
Let $N'(T_d)$ denote the total number of products
sold in insurance duration time.
It is easy to see that $N'(T_d)$
follows a Poisson distribution
with parameter $\lambda_2 P_{br} T_d$.
The worst case is that all buyers hold the
product till the
last minute of the clearing time $T_c$
and then return it.
Using a Chernoof bound \cite{Mitzenmacher2005} argument,
one can easily bound the shipment cost (at the worst case)
as
\[
\Pr[N'(T_d) C_S \geq N' C_S] \leq e^{-\lambda_2 P_{br} T_d}
(e \lambda_2 P_{br} T_d)^{N'} / N'^{N'}
\]
Setting $N' =\max\{\ln \epsilon^{-1} - \lambda_2 P_{br} T_d,
e^2 \lambda_2 P_{br} T_d\}$ we have
$\Pr[N'(T_d) C_S \geq N' C_S] \leq \epsilon$.
The clearing time follows the shipment delay $d$.
\done

\end{document}